\documentclass{article}[12pt]

\usepackage{amsfonts}
\usepackage{amsmath}
\usepackage{amssymb}

\begin{document}

\title{Dynamical curvature in a nonstandard cosmological model}
\author{Peter C. Stichel\\
Fakult\"at f\"ur Physik, Universit\"at Bielefeld\\
D-33501 Bielefeld, Germany\\ 
e-mail: peter@physik.uni-bielefeld.de}
\date{28.11.2017}

\maketitle

\begin{abstract}
We consider a nonrelativistic cosmological model introduced in [1] and derived as the nonrelativistic limit (or approximation at sub-Hubble scales) of a general relativistic model in [3, 4]. The latter is defined by an energy-momentum tensor containing only dust and a nontrivial energy flow. The nonrelativistic limit contains in leading order a 1st-order relativistic contribution to the spatial curvature whose time-dependence drives the accelerated expansion of the Universe (we do not need any kind of dark energy). Analytic solutions of the model are fixed by three constants (initial conditions). In the present paper we use our model as a toy model by adjusting the three constants in two different ways to a second order polynomial fit by Montenari and R\"as\"anen [5] to the observed expansion rate $H(z)$ for  
$z \lesssim 2$ (mainly cosmic chronometer data).  In scenario 1 we adjust our model to this fit and its derivative at the self-consistently determined transition redshift $z_t$.  In scenario 2 we use the same fit at $z_t$   and in addition $H(z)$ at decoupling  $(z = 1089)$. The Hubble parameter $H_0$   is taken from the polynomial fit in [5]: $H_0   = 64.2 km/s/Mpc$. For both scenarios we obtain a satisfactory agreement between predicted and observed $H(z)$ values. But the outcomes for the curvature function $k(z)$ are completely different:  In scenario 1 we obtain a strong variation of $k(z)$ ranging from $k(0) = - 1.216$ up to $k(2.33) = 0.718$. On the other hand scenario 2 shows an almost constant value for $k(z) \sim - 1$ for all $z \lesssim 2$ in agreement with the polynomial fit to one of the FRW consistency conditions performed in [5].
\end{abstract}

\section{Introduction}
It is beyond any doubt that the present Universe undergoes a phase of (real or apparent) accelerated expansion (cp. [6] and the literature cited therein).  Almost all observations are in good agreement with the standard cosmological model, however some observations are in disagreement with this $\Lambda$CDM model (see [7]).

Two alternative strategies to the $\Lambda$CDM model are under discussion.

In the first category one introduces some kind of ''new physics'' by changing Einstein's field equations (EFEs) either by modifying the geometrical part of the EFEs (called modified gravity), or by changing the matter part by adding some scalar, vector or tensor fields.

In the second category one considers accelerated expansion as an apparent effect due to averaging over inhomogeneities in the Universe (called backreaction, see [8] for a recent review).

For cosmological models based on averaging over inhomogeneities, one comes to the conclusion that the present day cosmic acceleration is due to a negative spatial curvature [9] (see also [10], [11], [12]).  A comparison of such backreaction effects with observations has been undertaken in [13].  Furthermore numerical solutions of Einstein's field equations for a Silent Universe show ''that the spatial curvature emerges due to nonlinear evolution of cosmic structures'' [14].

Whether a nontrivial spatial curvature exists can be tested by means of FRW consistency conditions (see [5]). One of them relates the FRW curvature parameter $k_H$ to the dimensionless expansion rate $h(z) = H(z)/H_o$ and the dimensionless comoving angular diameter distance $d(z) = (1+z)d_A   (z) = d_L   (z)/(1+z)$  ($d_L$ is the luminosity distance) [5]:
\begin{equation}
k_H(z) \equiv \frac{1-h^2d^{\prime 2}}{d^2}
\end{equation}

\noindent                                                                                                                                                       
Any z-dependence of $k_H$ would be a sign for violation of the FRW model.

In a very recent paper Montenari and R\"as\"anen have tested (1) by means of polynomial fits to observed values  of the expansion rate (second order polynomial) and the luminosity distance $d_L (z)$ (fourth order polynomial). It turns out that $k_H$    is almost constant for $z \lesssim 1.4$ (see fig. 6b in [5]; for any details we refer to [5]).

In the present paper we will show that a nonrelativistic cosmological model introduced in [1] and derived as the nonrelativistic limit (approximation at sub-Hubble scales) of a general relativistic model in [3, 4] is able to mimic such a behavior.  By an appropriate choice of three integration constants (see scenario 2 below) we obtain for the curvature function the prediction $k(z)  \sim - 1$ for $z \lesssim 2$.

The paper is organized as follows: In section 2 we review the essentials of our cosmological model. Two different ways (scenario 1 and scenario 2) of adjusting the three integrations constants of the model to a polynomial fit by Montenari and R\"as\"anen [5] to $H(z)$ data are described in section 3. In section 4 we summarize the numerical results and conclude in section 5 with some final remarks.

\section{Cosmological model}

We consider a self-gravitating fluid  (velocity field  $u_\mu$) with the following properties:
\begin{description}
\item {--} The fluid flow is geodesic  (vanishing acceleration).                                                   
\item {--} The fluid flow is irrotational.
\item {--} The energy-momentum tensor (EMT) $T_{\mu \nu}$, which represents the right hand side of Einstein's field equations (EFEs) ($\kappa \equiv 8\pi G$; $G_{\mu \nu}$ denotes the Einstein tensor)
\end{description}
\begin{equation}
G_{\mu\nu} = c^{-4} \kappa T_{\mu\nu}
\end{equation}
is supposed to be pressure-less and stress-free. Therefore the EMT is decomposed as
\begin{equation}
T_{\mu\nu} = \rho u_\mu u_\nu + q_\mu u_\nu + q_\nu u_\mu
\end{equation}
where $\rho$ is the energy density in the comoving frame and  $q_\mu$  is the energy flow 
vector ($u^\mu q_\mu = 0$.)  
                              ).  
The energy density $\rho$ consists of the dark sector contribution as well as the baryonic part.

In the nonrelativistic, shear-free limit (or at sub-Hubble scales) we obtain the following system of three coupled ordinary differential equations for the cosmological scale factor $a(t)$, the active gravitational mass density  $\rho (t)$ (we define $\hat{\rho} \equiv \kappa a^3 \rho/6$ ) and the energy flow vector  $q_i = q(t) x_i$ [3, 4]
\begin{equation}
\ddot{a} = - \frac{\hat{\rho}}{a^2} \ ,
\end{equation}             
and
\begin{equation}
\dot{\hat{\rho}} + \frac{\kappa}{2} q a^3 = 0
\end{equation} 
and
\begin{equation}
\dot{q} + 5 \frac{\dot{a}}{a} q = 0
\end{equation}
with  the curvature function $K(t) \equiv \frac{a^2}{6} R^{(3)}$ ($R^{(3)}$ is the spatial curvature) given by the Hamiltonian constraint
\begin{equation}
K = - \dot{a}^2 + \frac{2\hat{\rho}}{a} \ .
\end{equation}
We eliminate the function $q(t)$ by integrating (6)                                                                                                                                                                     
\begin{equation}
q(t) = \frac{2 K_1}{\kappa a^5 (t)},
\end{equation}
where the constant $K_1$ is determined by the initial value of $q$. 
    
The resulting dynamical system possesses two constants of motion $Q_2$ and  $Q_3$  [2,3]
\begin{equation}
Q_2 \equiv K_1 \dot{a} - \frac{1}{2} \hat{\rho}^2\ , ~~~Q_3 \equiv - \frac{\hat{\rho}^3}{6} - Q_2 \hat{\rho}  + \frac{K^2_1}{a}
\end{equation}
On the solution space of (4)-(6) the  $Q_i$ ($i = 2, 3$) take constant values $K_i$ which are determined by the initial values of $\rho$ and $H$.
  
After all we get from (9) energy density  $\rho$ and expansion rate $H \equiv \frac{\dot{a}}{a}$ as functions of the redshift $z$ ($1 + z \equiv a^{-1}$).  In dimensionless units
\begin{equation}
k_1 \equiv \frac{K_1}{H_0^3}\ , \qquad k_2 \equiv \frac{K_2}{H_0^4}\ , \qquad k_3 \equiv \frac{K_3}{H_0^6}\ ,
\end{equation}
$$
h(z) \equiv \frac{H(z)}{H_0}\ , \qquad \tilde{\rho} \equiv \frac{\hat{\rho}}{H_0^2}
$$
we obtain the following system of two coupled algebraic equations for $\tilde{\rho} (z)$ and $h(z)$ 
\begin{equation}
\frac{\tilde{\rho}^3 (z)}{6} + k_2 \tilde{\rho} (z) = k^2_1 (1+z) - k_3
\end{equation}                                                                                                                                                
and
\begin{equation}
h(z) = \frac{1+z}{k_1} \left(k_2 + \frac{1}{2} \tilde{\rho}^2 (z)\right)\ .
\end{equation}

Supposed the $k_i$ ($i = 1, 2, 3$) take positive values. Then, with the definition of the transition redshift $z_t$
\begin{equation}
1 + z_t \equiv k_3/k^2_1
\end{equation}
our cosmological equations (4), (11) and (12) describe for $z >  z_t ~(\rho  > 0)$ a decelerating phase and for $z <  z_t ~(\rho < 0)$ an accelerating phase of the Universe [1, 2].

Our primary interest is to obtain explicit expressions for the transition rate $h(z)$ and the dimensionless curvature function $k(z) \equiv K(z)/H^2_o$. To achieve this we
\begin{description}
\item{$\bullet$} square eq. (11) and get by means of (12) a cubic equation for $h(z)$
\begin{equation}
(k^2_1 (1+z) - k_3)^2 = \frac{2}{9} \left( \frac{k_1 h (z)}{1+z} - k_2\right) \left(\frac{k_1 h(z)}{1+z} + 2 k_2\right)^2
\end{equation}
and
\item{$\bullet$} insert (12) into (7) and get $k(z)$ in terms of $h(z)$
\begin{equation}
k (z) = - \left(\frac{h(z)}{1+z}\right)^2 \pm 2^{3/2} (1+z)  \left( \frac{k_1 h(z)}{1+z} - k_2\right)^{\frac{1}{2}}
\end{equation}
with the $+$ sign for $z > z_t$ and the $-$ sign for $z < z_t$.
\end{description}

For $z = 0$ eq. (14) leads to a constraint between the three $k_i$
\begin{equation}
(k_3 - k^2_1)^2 = \frac{2}{9} (k_1 - k_2) (k_1 + 2 k_2)^2 \ .
\end{equation}

\section{Fixing the integration constants by observations }

We have to fix the values of two independent constants $k_i$   and the value of the Hubble parameter $H_0$ by means of some data for the expansion rate $H(z)$. The only data for $H(z)$ which are independent of any cosmological model are the cosmic chronometer data (cp. table 1 in [5]). But these data possess still rather large errors. Furthermore our cosmological model is a nonrelativistic  model, its validity at larger redshifts is doubtful. So, a least-squares fit of these three constants to cosmic chronometer data seems not to be the very best. Instead we use our model as a toy model. We will show that two different adjustments of these constants to data for $H(z)$ lead both to a satisfactory agreement between predictions and observations. But the outcomes for the curvature function $k(z)$ become very different in the two scenarios.
 
For both scenarios we use the following second order polynomial fit by Montenari and R\"as\"anen to the $H(z)$ data (table 1 of [5]) [15]
\begin{equation}
h(z)  = h_1 z + h_2 z^2 \qquad         \mbox{with}~~ h_1 = 0.8368 \qquad \mbox{and} ~~    h_2   =  0.1082
\end{equation}
where we have listed only the mean values for the coefficients $h_{1,2}$.
 
For the Hubble parameter we take $H_0 = 64.2 km/s/Mpc$ determined by the same fit [5].
                                                                                                                     
\subsection{Scenario 1}

We determine the three constants $k_i$    by adjusting $h(z)$ and $h^\prime(z)$ at the self-consistently determined transition redshift $z_t$ to the polynomial fit (17).  We proceed in three steps:
\begin{description}
\item{$\bullet$} From the definition of the transition redshift  $\ddot{a}\mid_{z=z_t} =0$ we obtain, model independent
\begin{equation}
h^\prime (z_t) = \frac{h(z_t)}{1+z_t} \ .
\end{equation}
Using (17) in (18) we get
\begin{equation}
z_t = 0.58377\ .
\end{equation}  
\item{$\ast$} Eq. (12) taken at $z = z_t$ leads by means of (17) and (19) to
\begin{equation}
\frac{k_2}{k_1} = \frac{h(z_t)}{1+z_t} = 0.96313\ .
\end{equation}
\item{$\bullet$} The constraint eq. (16) may be rewritten as
\begin{equation}
k_1 = \frac{2}{9} z_t^{-2} \left( 1 - \frac{k_2}{k_1}\right) \left( 1 + 2 \frac{k_2}{k_1}\right)^2
\end{equation}
leading by means of (19) and (20) to
$$
k_1 = 0.20589
$$
and subsequently by means of (20) and (13) to
$$
k_2 =  0.19830  \qquad  \mbox{and}  \qquad    k_3    =  0.06714\ .
$$
\end{description}
Numerical results for the expansion rate $H(z)$ and the curvature function $k(z)$ for these values of the constants $k_i$ are given in table 2, section 4.

\subsection{Scenario 2}

In scenario 1 we have used besides the polynomial fit for $h(z)$ also its derivative. But the derivative of a fitting function is less reliable then the function itself.  So, in scenario 2 we use instead of $h^\prime(z)$ the value of the expansion rate at decoupling which, to a large extent, is independent of the late time cosmology [16] ($c$ denotes the velocity of light)
\begin{equation}
H(z = 1089)/c   = 5.2 /Mpc   \qquad       \mbox{or}  \qquad       H(1089)   =     15.6 \times 10^5      km/s/Mpc \ .
\end{equation}
The asymptotic behavior of $h(z)$ for $z\gg 1$ is determined, according to (14), only by the constant $k_1$
\begin{equation}
h(z) \sim \frac{6^{2/3}}{2} k_1^{1/3} z^{5/3} \ .
\end{equation}
So, for $z = 1089$ and by using (22) and $H_0   = 64.2 km/s/Mpc$ we obtain from (23)
\begin{equation}
k_1    = 0.002082 \ .
\end{equation}
Furthermore we use again the polynomial fit (17) at the self-consistently determined transition redshift $z_t$.  Then we obtain from (20, 21) a coupled system of two algebraic equations for $x = z_t$ and $y = k_2/k_1$
\begin{equation}
y = \frac{1 + h_1 x + h_2 x^2}{1 + x}
\end{equation}
and
\begin{equation}
x^2 = \frac{2}{9} k^{-1}_1 (1-y) (1 + 2y)^2\ .
\end{equation}
With (24) we obtain the solution
\begin{equation}
x = z_t = 1.473 \qquad        \mbox{and}     \qquad      y = \frac{k_2}{k_1} = 0.99773 \ .
\end{equation}
Finally by using (13) we get for the remaining constants $k_{2,3}$         
\begin{equation}
k_2   = 0.0020773 \qquad   \mbox{and} \qquad    k_3      = 1.072 \times 10^{-5}  \ .
\end{equation}
The values for the three constants $k_i$     given in (24) and (28) are very different from those obtained in scenario 1 (see subsection 3.1). Nevertheless, the corresponding values for the expansion rate $H(z)$ are as well in satisfactory agreement with the observed values (see table 2, section 4). But the predictions for the curvature function $k(z)$ are completely different. We obtain almost constant values $k(z) \sim - 1$ for all $z \lesssim 2$ (see table 2, section 4).

\newpage

\section{Results and conclusions}
 
In this section we will summarize our results. In table 1 we specify the constants $k_i $ for both scenarios.  In table 2 we list the corresponding predictions for the expansion rate $H(z)$ and the curvature function $k(z)$ compared to data for $H(z)$ and to the polynomial fit (17).

\bigskip
\bigskip
\noindent
\begin{tabular}{l | l | l  | l}
\hline
\hline                      &         $k_1$                   &  $k_2$                            &           $k_3$ \\ \hline
Scenario 1               &    0.20589                  &         0.19830               &               0.06714\\
Scenario 2               &      0.0020820              &         0.0020773             &           1.072 x $10^{-5}$\\
\hline
\hline
\end{tabular}	 

\medskip
\noindent
{\bf Table 1:} The constants $k_i$  for scenario 1 and scenario 2, determined in section 3.

\bigskip
\bigskip
\noindent
\begin{tabular}{c c}
\hline\hline
~~~~~~~~~~~~~~~~~~~~~~~~~~~~~~~~~~~~~~~~~~~~~~~~~~~   Scenario 1     ~~~~~~~  &     Scenario 2~~~~~~~~~\\  
\end{tabular}

\noindent
\begin{tabular}{l | l | r | r | r | c | r }
\hline                                                                                 
  z     &      ~~~~~$H_{ob}(z)$        &    $H_{pol}(z)$   &        $ H(z)$   &      $k(z)$  &                 $H(z)$     &           $k(z)$\\ \hline

0.07    &      ~~~~69~+ 19.6          &       67.995    &             68.134   &        - 1.22    &             68.679  &      - 1.00582\\
0.12    &      ~~68.6~+ 26.2        &       70.747    &             70.941   &      - 1.19      &          71.880  &      - 1.00559\\
0.179   &      ~~~~75~+ 4        &       74.039    &             74.260   &        - 1.17    &          75.653  &     - 1.00531\\
0.199   &      ~~~~75~+ 5         &       75.166    &             75.388   &         - 1.16   &          76.932  &     - 1.00521\\
0.2     &      ~~72.9~+ 29.6     &       75.222    &             75.445   &        - 1.15    &          76.996  &    - 1.00520\\
0.28    &      ~~88.8~+ 36.6      &       79.787    &             79.982   &        - 1.11    &          82.112  &    - 1.00478\\
0.32    &       ~(78.6 + 27)*      &       82.103    &             82.270   &        - 1.09    &          84.669  &   - 1.00456\\
0.352   &      ~~~~~~8~+ 14~~        &       83.971    &             84.113   &         - 1.07   &          86.715  &    - 1.00438\\
0.3802  &      ~~~~~83~+ 13.5        &       85.629    &             85.748   &        - 1.06    &          88.518  &    - 1.00422\\
0.4004  &      ~~~~~77~+ 10.2       &       86.824    &             86.925   &        - 1.04    &          89.810  &    - 1.00410\\
0.4247  &      ~~~87.1~+ 11.2        &       88.269    &             88.349   &        - 1.03    &          91.363  &    - 1.00396\\
0.4497  &      ~~~92.8~+ 12.9        &       89.764    &             89.824   &        - 1.01    &          92.960  &    - 1.00381\\
0.4783  &      ~~~80.9~+ 9.0         &       91.485    &             91.524   &        - 1.00    &          94.789  &    - 1.00363\\
0.48    &      ~~~~~97 + 62      &       91.587    &             91.626   &        - 1.00    &          94.898  &    - 1.00362\\
0.57    &        ~(96.9 + 2.8)*    &       97.079    &            97.080    &       - 0.94     &          100.651 &    - 1.00304\\
0.593   &       ~~~104~+ 13         &       98.500    &             98.501   &        - 0.92    &          102.121 &     - 1.00289\\
0.680   &       ~~~~92~+ 8        &       103.943   &             103.990  &        - 0.86    &          107.682 &     - 1.00228\\
0.781   &        ~~~105~+ 12        &       110.394   &            110.616   &       - 0.79     &          114.130 &     - 1.00154\\
0.875   &        ~~~125~+ 17        &       116.526   &            117.062   &       - 0.71     &          120.147 &     - 1.00081\\
0.88    &        ~~~~90~+ 40      &       116.855   &             117.413  &        - 0.71    &          120.466 &     - 1.00077\\
1.037   &        ~~~154~+ 20        &       127.380   &            128.889   &       - 0.58     &          130.505 &     - 0.99945\\
1.363   &        ~~~160~+ 33.6      &       150.329   &            155.892   &       - 0.29     &          151.362 &     - 0.99559\\
1.965   &      186.5~+ 50.4       &       196.587   &          218.891     &    + 0.32        &          189.969 &     - 0.99004\\
2.33    &        ~(224 + 8)*       &       227.085   &             265.992  &          + 0.72  &          213.465 &    - 0.98519\\
\hline\hline
\end{tabular}

\medskip
\noindent
{\bf Table 2:} Expansion rate data $H_{ob}$ (2nd column, taken from table 1 in [5]; data marked with a * are data, all others are cosmic chronometer data)) versus polynomial fit $H_{pol}$      [5](3rd column; see eq. (17)) and predictions for $H(z)$ from scenario 1 (4th column) and scenario 2 (6th column). For the polynomial fit as well as for the predictions we have used $H_0 = 64.2 km/s/Mpc$  [5].  The 5th and the 7th column show the predictions for the curvature function $k(z)$ for scenario 1 and scenario 2.  

\subsection*{Comments on the results for scenario 1:}

The predictions for the expansion rate $H(z)$ (see table 2) are well behaved for $z \lesssim 1.363$ but they become too large for $z \gtrsim 2$. The reason for this is presumably the nonrelatistic nature of our model. Indeed we expect relativistic corrections for larger z-values. 

The curvature function $k(z)$  shows a strong variation with redshift $z$ and a transition from a hyperbolic to a spherical space at a z- value between z = 1.363 and z = 1.965.

\subsection*{Comments on the results for scenario 2:}

The predictions for the expansion rate $H(z)$ (see table 2) are systematically higher than those of scenario $1$ for  $z \lesssim 1$, but they are still good-looking. For higher z-values the agreement with the data becomes better. This is no surprise in the light of our adjustment to $H(z)$ data at decoupling.

But the predictions for the curvature function $k(z)$ are completely different from those of scenario 1. They show an almost constant behavior $k(z) \sim - 1$ for all $z \lesssim 2$ in agreement with the polynomial fit to the FRW consistency condition (1) [5]. This almost constant behavior follows from the small value for $k_1$    (24), as the slope of $k(z)$ is proportional to $k_1$(take the time derivative of (7) and use (4), (5) and (8))
\begin{equation}
k^\prime (z) = \frac{2 k_1 (1+z)^2}{h(z)}\ .
\end{equation}
Asymptotically  ($z \gg 1$) we obtain from (15) and (23)
\begin{equation}
k (z) \sim \frac{6^{1/3}}{2} k_1^{2/3} z^{4/3}
\end{equation}
leading at decoupling to
\begin{equation}
k(z = 1089)     \sim       166
\end{equation}
So, at decoupling, we have a spherical space for both scenarios.

\section{Final remarks}

Our results show that the observation of an almost constant curvature function $k(z)$    from the FRW consistency condition (1) [5] is not automatically an evidence for the validity of the FRW model. Of course, the FRW model is one possible model in this case, but it is not the only one. Using our nonrelativistic cosmological model as a toy model  by adjusting three free constants (initial conditions) to a polynomial fit by Montenari and R\"as\"anen to $H(z)$ data for  $z \lesssim 2$ and to the expansion rate at decoupling our model generates a dynamically determined curvature function which is almost constant for  $z \lesssim 2$. Certainly this does not proof that our model describes the mechanism which produces such a behavior. But it has been shown that another mechanism than the FRW model, where a constant curvature $k
$ is an ad hoc parameter, is possible.

 To answer the question, whether our cosmological model can be more than a toy model we should  progress in two respects:
\begin{description}
\item{$\bullet$}	 We should either provide a cosmological solution of the general relativistic equations (2, 3) or, at least, evaluate some relativistic corrections to the present nonrelativistic model.
\item{$\bullet$}	Our model should pass more cosmological tests. Besides expansion rate and curvature function we have until now only the stationary solution of equations (2, 3) described by one nonlinear ordinary differential equation for the gravitational potential [17] whose weak field limit [4] describes successfully galactic halos [2]. 
\end{description}

\section*{Acknowledgements}

I'm grateful to Thomas Buchert for discussions and valuable hints. I thank Francesco Montenari for providing the polynomial fits from [5].

\newpage

\section*{References}

\begin{description}
\item{[1]} P.C. Stichel and W.J. Zakrzewski, Can cosmic acceleration be caused by exotic massless particles, 
      Phys. Rev. D 80, 083513 (2009).

\item{[2]} P.C. Stichel and W.J. Zakrzewski, Nonstandard approach to Gravity for the Dark Sector of the 
     Universe,  Entropy 15, 559 (2013).

\item{[3]} P.C. Stichel, Cosmological model with dynamical curvature, arXiv:1601.07030.

\item{[4]} P.C. Stichel and W.J. Zakrzewski, General relativistic, nonstandard model for the dark sector of the 
     Universe, Eur. Phys .J. C, 75:9 (2015).

\item{[5]} F. Montenari and S. R\"as\"anen, Backreaction and FRW consistency conditions, arXiv:1709.06022. 

\item{[6]} D. Huterer and D.L. Shafer, Dark energy two decades after: Observables, probes, consistency 
      tests, arXiv:1709.01091.

\item{[7]} T. Buchert et al., Observational Challenges for the standard FLRW Model,
       Int. J. Mod. Phys. D 25, 1630007 (2016)
       
\item{[8]} T. Buchert et al., Is there proof that backreaction of inhomogeneities is irrevelant in cosmology? Class. Quant. Grav. 32, 215021  (2015)     
       
\item{[9]} T. Buchert, Dark Energy from structure: a status report, Gen. Rel. Grav. 40, 467 (2008).

\item{[10]} T. Buchert and M. Carfora,  On the curvature of the present-day Universe, Class. Quant. Grav. 25, 195001 (2008).

\item{[11]} K. Bolejko, Emergence of spatial curvature, arXiv:1707.01800

\item{[12]} K. Bolejko, Cosmological backreaction within the Szekeres model and emergence of spatial curvature, JCAP 06, 025 (2017)

\item{[13]} J. Larena et al., Testing backreaction effects with observations, Phys. Rev. D 79, 083011 (2009)

\item{[14]}  K. Bolejko, Towards concordance cosmology with emerging spatial curvature, arXiv:1708.09143.

\item{[15]} F. Montenari, private communication.

\item{[16]} L. Verde et al. Early Cosmology Constrained,  JCAP, 023, 1704 (2017).

\item{[17]} P. Stichel, Anisotropic geodesic fluid in non-comoving spherical coordinates, arXiv:1706.02982

\end{description}
\end{document}